\documentclass[a4paper,12pt]{article}
\usepackage[utf8]{inputenc}
\usepackage[T1]{fontenc}
\usepackage[swedish,english]{babel}
\usepackage{color}
\usepackage{amssymb}
\usepackage{amsmath}
\usepackage{ae}
\usepackage{slashed}
\usepackage{graphics}
\usepackage{graphicx}
\usepackage{epstopdf}
\usepackage{url}

\newcommand{\g}{\gamma}

\newcommand{\lb}{\bar{\lambda}}
\newcommand{\la}{\lambda}

\newcommand{\lbracket}{[\![}
\newcommand{\rbracket}{]\!]}

\DeclareMathOperator{\tr}{tr}
\usepackage{cite}
\usepackage{epsfig}
\usepackage{times}
\usepackage[small]{caption}
\usepackage{ifthen}
\usepackage{longtable}
\usepackage{multirow}
\numberwithin{equation}{section}

\usepackage{fancyhdr}
\usepackage[colorlinks=true,linktoc=page,linkcolor=blue,citecolor=blue,filecolor=blue,urlcolor=blue]{hyperref}

\begin{document}
\selectlanguage{english}
\setcounter{secnumdepth}{3}
\frenchspacing
\pagenumbering{roman}

\null{\vspace{\stretch{1}}}
\includegraphics*[width=1.8cm]{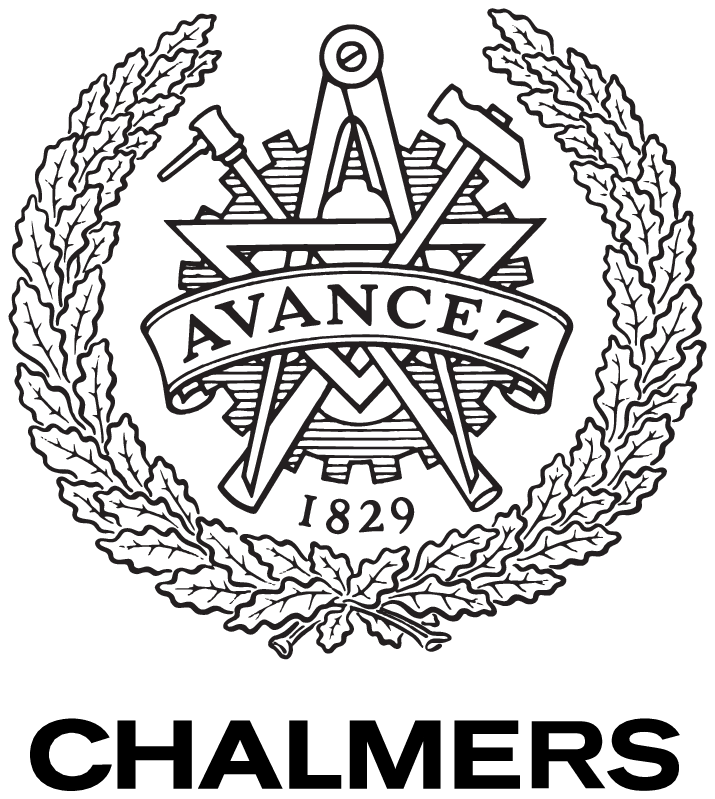}
\hfill $\begin{array}{r}\text{Gothenburg} \\ \text{August,
    2015}\\$\quad$\end{array}$
\vskip-3mm
\noindent\makebox[\linewidth]{\rule{\textwidth}{0.4pt}}
\vspace{\stretch{1}}
\begin{center}
{\Large Ultraviolet divergences in maximal supergravity\\ \vspace{0.1cm} from a pure spinor point of view}\\
\vspace{\stretch{1}}
{\normalsize Anna Karlsson}\\
\vspace{\stretch{1}}
{\small Fundamental Physics \\ Chalmers University of Technology \\ SE 412 96 Gothenburg, Sweden }\\
\vspace{\stretch{1}}
\end{center}

\begin{abstract}\noindent
The ultraviolet divergences of amplitude diagrams in maximal supergravity are investigated using the pure spinor superfield formalism in maximal supergravity, with maximally supersymmetric Yang--Mills theory for reference. We comment on the effects of the loop regularisation in relation to the actual absence of high powers (within the degrees of freedom) of the non-minimal variable $r$. The absence affects previous results of the field theory description, which is examined more closely (with a new $b$-ghost) with respect to the limit on the dimension for finiteness of the theory, dependent on the number of loops present. The results imply a cut-off of the loop dependence at six loops for the 4-point amplitude, and at seven loops otherwise.
\end{abstract}
\vspace{\stretch{2}}
\vfill
\noindent\makebox[\linewidth]{\rule{\textwidth}{0.4pt}}
{\tiny \texttt{email: karann@chalmers.se}}
\thispagestyle{empty}
\newpage

\noindent\makebox[\linewidth]{\rule{\textwidth}{0.4pt}}
\vspace{-0.8cm}
\tableofcontents
\vspace{0.3cm}
\noindent\makebox[\linewidth]{\rule{\textwidth}{0.4pt}}
\pagenumbering{arabic}

\section{Introduction}
While the ultraviolet divergences in maximally supersymmetric Yang--Mills theory \cite{BSS} are well known \cite{M,BLN,HST} the case of maximal supergravity \cite{SA,Brink:1980az,Cremmer:1980ru,Wit:1982ig} remains an open question. Explicit calculations for the four-graviton amplitude diagram in four dimensions ($\mathcal{N}=8$, four external states) have reached four loops \cite{Bern2,Bern25,Bern:2012uf}, up until which point the theory remains finite under the same conditions on the dimension as for super-Yang--Mills theory. Investigations through more general methods \cite{Grisaru:1976ua,Grisaru:1976nn,Deser:1977nt,Tomboulis:1977wd,Deser:1978br,Deser:1979sx,Berkovits:2004px,Berkovits:2006vi,vanB,KLR,BHS,AB,BD,Vanhove:2010nf,Bossard:2010dq,Beisert:2010jx,BG,JB,CK2} on the other hand, are not conclusive on what conditions will hold for maximal supergravity. Some studies point at a first possible divergence at seven loops in four dimensions, for the 4-point amplitude \cite{Vanhove:2010nf,BG,JB,Beisert:2010jx,CK2}. In fact, a scenario with maximal supergravity as finite in four dimensions would be slightly confusing, as the theory then would present a well-defined quantum theory, possible to treat perturbatively without any alteration. String theory or M-theory is expected to present its ultraviolet completion, and the general discussion (in short) concerns at what loop order the theory diverges.

In this, a few investigations using the pure spinor formalism \cite{Berkovits:2004px,Berkovits:2006vi,AB,BG,JB,CK2} have been performed. The advantage of the formalism is that it makes it possible to keep the maximal supersymmetry manifest off-shell: there exist actions for both maximally supersymmetric Yang--Mills theory and maximal supergravity. As such, desired simplifications might be expected to occur. Of the studies previously mentioned, the first ones were initiated by Berkovits \cite{Berkovits:2004px,Berkovits:2006vi,AB} and later partly extended by Bj\"ornsson and Green \cite{BG,JB}. They are performed from a string theory point of view, with the examinations of the amplitudes chiefly taking place in maximally supersymmetric Yang--Mills theory, before the results are generalised to maximal supergravity. As a result, the maximal supersymmetry is not manifestly present throughout the examinations, though the end result is characterised by it. Our previous paper \cite{CK2} instead started out from the supergravity action, constructing a field theory description for the amplitudes. In this way, the description could benefit from manifest, maximal supersymmetry throughout the examination.

The results of \cite{CK2} were in accordance with those obtained by the previous investigations using pure spinors. The predicted limits for UV finiteness were in agreement with each other, as were the limitations of the evaluation methods. However, we have since approached the subject in a more detailed manner, the result which is the subject of this article. 

Our findings are peculiar.\footnote{In our original analysis, we mistakenly ignored the possible contribution of nonzero modes of $N_{mn}$ in the $b$-ghost. It is possible that contributions from these nonzero modes will modify our conclusions, and we are currently investigating this. We thank Nathan Berkovits for pointing this out.} We argue that the previous investigations using pure spinors have not taken the full loop regularisation into account, not as far as is necessary, which is why we no longer can provide results in unison with what is known from the explicit calculations for the four-graviton amplitudes: our method is not predictive enough, and we can only state that the 4-point amplitude is finite in three dimensions, and possibly for higher dimensions as well. Our previous conclusion that something new might be expected to occur at five loops (if ever) does not necessarily hold. Either algebraic relations not yet discerned by us occur already at two loops, or at different stages at and above two loops. 

To begin with, we take a closer look at the loop regularisation introduced in order to allow for loop momenta to exist in the amplitude description, while retaining the possibility to perform examinations of the behaviour of the amplitudes. The most important finding is that the operators in the loop structures cannot provide $r^{11}$ (i.e. $r$ to its degrees of freedom) in maximally supersymmetric Yang--Mills theory, nor $r$ (of $23$ degrees of freedom) to any higher power than $15$ in maximal supergravity. This affects how far the loop regularisation must be performed (to a greater extent than previously assumed). Another key feature related to the loop integrations is the presence of total derivatives (i.e. vanishing terms).  

Secondly, we have reformulated the $b$-ghost to simplify the analysis of how terms in the amplitudes can combine in order to satisfy the loop integrations brought on by the loop regularisation, while creating the divergences in the ultraviolet regime. 

The most intriguing observation is that the amplitudes behave no worse than at seven loops, and the 4-point amplitude no worse than at six loops, as one-particle irreducible loop structures above those limits cannot be formed. In a worst case scenario, in combination with the results of \cite{Vanhove:2010nf,BG,JB,Beisert:2010jx}, this would mean that the theory of maximal supergravity in four dimensions diverges merely logarithmically.

The article is organised as follows. We begin by giving a brief recapture of the essential concepts of the field theory description of the amplitudes in the pure spinor formalism, and refer to \cite{CK2} for a detailed description\footnote{An overview of the pure spinor formalism exist in e.g. \cite{Cederwall:2013vba}.} (or \cite{Karlsson:2014xoa} in a briefer format). Here, we use the case of maximally supersymmetric Yang--Mills theory to illustrate the principles, as Yang--Mills provides a well-known theory and its field theory description is far more concise than what is true for supergravity. The limits on the dimension for finiteness in Yang--Mills theory may be observed in \eqref{eq.DSYM}.

We then proceed to the properties of the loop regularisation in relation to the divergences with respect to small $(\la,\lb)$ and the subsequent recognition of extra degrees of freedom of $r$. This occurs at an earlier stage than previously assumed, due to an effective limit on the power of $r$ below its degrees of freedom, discussed in connection to \eqref{eq.rL1015}.

Finally, we proceed to the case of maximal supergravity. Here, a new shape of the $b$-ghost, better fitted for calculations than the one in \cite{CK2}, is presented. We analyse the basic principles for how the components in the $b$-ghost and the operators in the vertices can combine to satisfy the loop integrations while furnishing the ultraviolet divergences. For a definite statement on the UV divergences though, further examinations are necessary.
Perhaps manifest U-duality is required, or properties not yet recognised play a vital r\^ole in the cancellation of certain terms.

The requirement of $L\leq6$ for 4-point one-particle irreducible loop structures is discussed prior to \eqref{eq.L6} and the general requirement of $L\leq 7$ concerning the $n$-point ($n\geq 4$) equivalents around \eqref{eq.L7}. Effectively, we observe a cut-off of the behaviour of the amplitudes at (six) seven loops. This occurs naturally for supersymmetric Yang--Mills theory in the sense that the limit on (whole) dimensions does not alter past six loops, but has interesting implications for supergravity.

\section{Amplitude diagrams in the pure spinor formalism}
The properties of the amplitude diagrams in $D=10$, $\mathcal{N}=1$ supersymmetric Yang--Mills theory and $D=11$, $\mathcal{N}=1$ supergravity are, in a field theory approach in the pure spinor formalism, set by the actions\cite{B1,B3,Cederwall:2009ez,Cederwall:2010tn}:
\begin{subequations}\label{eqs.actions}
\begin{align}
S_{SYM}&\sim\int [\mathrm{d}Z] {\tr}\bigg(\frac{1}{2}\psi Q\psi+\frac{1}{3}\psi^3\bigg)\label{eq.SYM},\\
S_{SUGRA}&\sim\int[\mathrm{d}Z]{\tr}\bigg(\frac{1}{2}\psi Q\psi+\frac{1}{6}(\la\g_{ab}\la)\Big(1-\frac{3}{2}T\psi\Big)\psi R^a\psi R^b\psi\bigg).\label{eq.SUGRA}
\end{align}
\end{subequations}
Each is characterised by manifest, maximal supersymmetry and reminiscent of a Chern--Simons action, with $Q$ being a BRST operator and $\psi$ the pure spinor superfield. The first part represents the abelian action, whereas the latter describes the interactions. As can be noted, only 3-point vertices exist in the Yang--Mills theory, whereas supergravity also contains 4-point vertices. In addition, there exists operators connected to the vertices in supergravity, $R^a$ and $T$, which by necessity act out of the vertices on different legs, the different configurations of which are equivalent. In Yang--Mills theory, the only truly relevant operator in the description, apart from $Q$, is the propagator.

\subsection{The pure spinor formalism, in short}
The actions in \eqref{eqs.actions} originate in an observation of the properties of the abelian theories: the covariant spinor derivative $D_\alpha$ acting on the 1-form $C_\alpha(x,\theta)$ (Yang--Mills theory) and the 3-form $C_{\alpha\beta\gamma}(x,\theta)$ (supergravity), each containing the physical fields of the theories, represent the equations of motion, contingent upon certain irreducible representations being removed. These properties are captured in full \cite{B1,B3,Brink:1980az,Cremmer:1980ru,Cederwall:2009ez} by a construction \cite{B1,B3} with the bosonic spinor $\la^\alpha$ (of ghost number one) contracted with the indices of the derivative and physical fields, provided the spinor is pure:
\begin{equation}
\la^\alpha:\quad \la\g_a\la=0.
\end{equation}
That is, $Q\psi=0$ with $Q=\la D$ etc. gives the equations of motion, with
\begin{equation}\label{eq.DDp}
\{D_\alpha,D_\beta\}=-2(\g^a)_{\alpha\beta}\partial_a.
\end{equation}
In addition, the components of $\psi$ remain untouched until subject to the equation of motion, which allows for off-shell degrees of freedom and an extension of the theory so that $\psi$ contains more fields (of non-zero ghost number) than $C$: antifields etc. In this way, the necessary components for a formulation with an action are introduced, while the original, free theory remains retainable at ghost number zero in $\psi$.

Moreover, \eqref{eq.DDp} shows $Q$ to be a BRST operator, and the description above that of a BRST formulation, with a natural extension to a theory of interactions in the Batalin--Vilkovisky formalism \cite{BV,BR}. In essence, the symmetry operator $Q$ is replaced with a generalised action acting on fields nonlinearly through an antibracket, the only available option (as it is not desirable to split $\psi$ into fields and antifields) being \cite{Cederwall:2009ez}
\begin{equation}
(A,B)\sim\int\frac{\delta A}{\delta\psi}\frac{\delta B}{\delta\psi}[\mathrm{d}Z],
\end{equation}
resulting in a formulation with the equation of motion $(S,\psi)=0$. The action is obtainable through the master equation $(S,S)=0$, with a starting point in the BRST action (the first parts of \eqref{eqs.actions}) for consistency. However, for a full description of the theories, further characteristics of the pure spinor formalism are of vital importance: BRST equivalence, gauge fixing, the concept of integration and that of general regularisations.

Firstly, BRST equivalence ($Q$-equivalence) represents a freedom of the formalism caused the fact that any calculation is performed between on-shell, external states. Consequently, any term is only defined up to BRST equivalent terms:
\begin{equation}
1 \leftrightarrow 1+\{Q,\chi\},
\end{equation}
$\chi$ being a fermion of appropriate ghost number and dimension. So called regulators, $e^{\{Q,\chi\}}$, can be introduced at any time.

Secondly, gauge fixing cannot be performed in the conventional way for the same reason as mentioned in connection to the antibracket. Instead, gauge fixing in string theory is imitated through a Siegel gauge \cite{Siegel} resulting in a propagator $b/p^2$ where the $b$-ghost obeys
\begin{equation}\label{eq.bReq}
\{Q,b\}=\partial^2,\quad b\psi_\text{on-shell}=0
\end{equation}
and automatically should have the property of $bb=0$.

Thirdly, to obtain an integral measure which captures the full dynamics, the superspace is typically extended \cite{B2} to contain two extra variables $(\lb_\alpha,r_\alpha)$, counterparts to $(\la^\alpha,\theta^\alpha)$ obeying $\lb\g^a\lb=\lb\g^ar=0$, with $\lb$ of ghost number $-1$. In this non-minimal formalism\footnote{The derivative with respect to $\la$ is denoted by $\omega$ and shows up in the gauge invariant ($\omega_\alpha$ modulo $(\g_a\la)_\alpha X^a$, $X$ any 1-form) configurations of $N=\la\omega$, $N^{ab}=\la\g^{ab}\omega$ \cite{B2}. The corresponding is true for $\lb$. For $r$ there is $s$: $S=\lb s$, $S^{ab}=\lb\g^{ab}s$. Our conventions for the derivatives in supergravity are $[\la_\alpha,\omega_\beta]=\varepsilon_{\alpha\beta}$ and $\{r_\alpha,s_\beta\}=\varepsilon_{\alpha\beta}$, as further commented on in \S\ref{par.msugra}.} $Q$ is extended to $Q=\la D+r\bar{\omega}$, a regulator with $-(r\theta+\la\lb)$ in the exponent can be constructed, and a well-defined concept of integration can be formulated. The regulator removes any divergences in the limit of large $(\la,\lb)$ and furnishes extra $(r,\theta)$ for the integration over the non-minimal superspace variables, should that be necessary. That is, if the amount provided by the operators in the amplitude diagrams does not suffice to satisfy e.g.
\begin{equation}\label{eq.dr}
[\mathrm{d}r]\sim\lb_{\alpha_1}\ldots\lb_{\alpha_i}\star{\bar{T}^{\alpha_1\ldots\alpha_i}}_{\hspace{1cm}\beta_1\ldots\beta_j}\frac{\partial}{\partial r_{\beta_1}}\ldots \frac{\partial}{\partial r_{\beta_j}},
\end{equation}
where the set $(i,j)$ is $(3,11)$ in Yang--Mills theory \cite{B2} and $(7,23)$ in supergravity \cite{Cederwall:2009ez,Berkovits:2002uc,Anguelova:2004pg}.

Finally, general regularisations are necessary in the non-minimal formalism as too high, negative powers of scalars\footnote{In maximally supersymmetric Yang--Mills theory the only scalar possible to form out of $(\la,\lb)$ is $\xi=(\la\lb)$. For finiteness, the power of it may not go below $-10$ \cite{B2}. In maximal supergravity there also exists $\eta=(\la\g^{ab}\la)(\lb\g_{ab}\lb)$, where each 2-form is referred to by $\sigma$. There, the power of $\xi$ may not go below $-22$ for finiteness, whereas the equivalent for $\sigma$ is $-11$ \cite{Cederwall:2009ez}.} of $(\la,\lb)$ otherwise would cause divergences with respect to small $(\la,\lb)$. This can always be avoided through the introduction of new sets of variables and corresponding regulators and integral measures \cite{Berkovits:2006vi,CK2}. The procedure has a definite drawback though: it renders examinations of the remaining expressions severely difficult. Effectively, it is not strictly put to use, though vital as a concept. However, at the introduction of loops the unregularised propagator is too local to allow for loop momenta of $(\la,\lb)$. The current, string theory inspired solution to this problem, the loop regularisation \cite{Berkovits:2006vi}, consists of recognising the momenta of the loop(s) as variables, in effect\footnote{For convenience, the loop components are kept implicit in the notation.}
\begin{equation}
\partial_a\rightarrow \partial_a+\sum_I \partial^I_a
\end{equation}
for each derivative on a propagator making up a part of the loop(s) $I$, and allowing for an integration over the loop momenta. The procedure comes with the introduction of a regulator with exponent:
\begin{equation}\label{eq.loopreg}
k\big((\la D)S+(\la\g_{ab} D) S^{ab}-N\bar{N}-N_{ab}\bar{N}^{ab}\big),\quad k>0.
\end{equation}
The propagator regularised in this way is not too local for loop integrations, while examinations of the amplitudes can be performed. The latter however includes the consideration of a conversion of the variables in the operators present in the loop(s), when acted on by \eqref{eq.loopreg}. Most notably, a conversion of
\begin{equation}\label{eq.rTransf}
e^{\{Q,\chi\}}r_\alpha e^{-\{Q,\chi\}}:\quad r_\alpha \rightarrow r_\alpha+k(\g_{ab}\lb)_\alpha(\la\g^{ab}D),\end{equation}
is allowed for\footnote{The conversion into $\lb_\alpha(\la D)$ renders a zero result by the irreducible representations of $(\lb,r)$ in the operators, which are displayed in \eqref{eq.bSYM} and \eqref{eq.oSUGRA}.}. The properties of this are discussed in more detail in \S\ref{par.loopreg}.

\subsection{The structure of the amplitude diagrams}
A general amplitude diagram may be divided into the following parts: external fields, tree diagrams and loop clusters, the last term denoting one-particle irreducible sets of loop diagrams. The ultraviolet behaviour of an amplitude is no worse than that of the most badly behaved loop cluster in it, which is why the focus of these examinations is on loop clusters of $L$ loops, connected to other diagram parts through $j$ outer legs. Moreover, for the purpose of discerning the UV divergences in maximally supersymmetric Yang--Mills theory and maximal supergravity, it is sufficient to consider diagrams constructed out of 3-point vertices, a fact that will become apparent in \S\ref{par.msugra}.

In addition, it can be shown\footnote{\S\ref{par.lUV} and \S\ref{par.csD} discuss this in more detail with respect to the pure spinor formalism.} that any loop must have at least four outer legs: there are no bubbles or triangles \cite{BjerrumBohr:2008ji,ArkaniHamed:2008gz}. As such a general loop cluster (planar or non-planar) consists of
\begin{subequations}\label{eq.comps}
\begin{equation}
3(L-1) + j
\end{equation}
propagators, some\footnote{For $L>1$, the number of outer propagators is $j$. $L=1$ constitutes a special case as one propagator must represent the inner propagator, yielding a total of $j-1$ outer propagators.} of which are outer (existing due to the outer legs) and
\begin{equation}
2(L-1)+j
\end{equation}
vertices, including $j$ outer ones. For loop clusters, the limits on outer legs are
\begin{equation}\label{eq.jLimGen}\begin{array}{ll}
j\geq4,\quad&L\leq2\\
j\geq3, &L=3\\
j\geq2, &L\geq4,
\end{array}\end{equation}
\end{subequations}
where $L<4$ appear as special cases due to the absence of bubbles and triangles and the overall limit of $j\geq2$ occurs for reasons concerning total derivatives and the non-existence of vacuum amplitudes.

For our analysis of the loop clusters, we will look at the parts provided by the propagators and vertices present and how they can combine: the number of them (as stated above) is sufficient for a full analysis, and will be used throughout this article. The different loop configurations are disregarded, but included in the general analysis.

\subsection{The loop regularisation \& the ultraviolet divergences}\label{par.lUV}
With the loop regularisation comes integrations over the loop momenta: $\partial$, $D$, $\omega$, $\bar{\omega}$ and $s$. The first one causes the ultraviolet divergences when too high a power of $\partial^2$ is present, formed out of the components of the diagram in combination with the loop regularisation as described in \eqref{eq.rTransf}. Components outside a loop cluster do not contribute with any loop momenta and therefore do not affect the ultraviolet divergences. Inside a loop cluster, momenta are shared between the loops and sometimes forced out of a loop or the loop cluster. In total, the condition on the theory in order for it to be finite in the ultraviolet regime is
\begin{equation}\label{eq.UVfinC}
LD-6(L-1)-2j+2m<0,
\end{equation}
with $L$ denoting the number of loops present in the loop cluster, $D$ the dimension (possibly after dimensional reduction) and $m$ the number of $\partial^2$ formed out of the operators and the regulators in the loop cluster. This occurs as each propagator carries $1/\partial^{2}$ and unpaired $\partial_a$:s remaining in a loop give a zero contribution, either at the integration $[\mathrm{d}p^I]$ or as a total derivative, which will be further discussed in \S\ref{par.charL}. If several loop clusters are present, the condition must hold for each of them.

The key issue in the investigations of the UV properties is the maximal $m$, dependent on the (regularised) operators in the loop cluster. The derivatives in these remain in a/their loop if possible, otherwise they act on other components: on other derivatives/variables or out of the loop. Out of the remaining ones, the integration over the loop variables must be satisfied. $[\mathrm{d}s^{I}]$ brings down the necessary $s^{23}$ from $\la\lb D s$ of the loop regulator. $[\mathrm{d}D^{I}]$ then claim $D^{32}$ before the remaining expression needs to be analysed with respect to $m$.

We will use maximally supersymmetric Yang--Mills theory as an illustrative example in this, before discussing the loop regularisation in \S\ref{par.loopreg} and the more complicated case of maximal supergravity in \S\ref{par.msugra}. In Yang--Mills theory, five original $D$:s, i.e. \emph{not from the loop regulator}\footnote{The loop regulator provides $\la\g^{(2)}D$ and $\la D$, out of which at most $D^{11}$ can be constructed. Moreover, the original $D$:s in $b$ can give at most $D^5$ (antisymmetrised covariant derivatives).}, are claimed by $[\mathrm{d}D^{I}]$ per loop. In addition, for reasons that will be clear in \S\ref{par.loopreg}, each\footnote{We will denote the parts of $b$ by their (unregularised) power of $r$, as $b_n$: $0\leq n\leq3$. While referring to a $b$-ghost on an outer leg, we will use $b_j$.} $b_j$ contributes with no more than one such $D$ (or $\partial$ in its stead) as $bb=0$ forces at least one derivative out of the loop cluster, a process treatable without loop regularisation, by $Q$-equivalence. This immediately prohibits bubbles and triangles, also noted in \cite{BG,JB}.

The $b$-ghost in maximally supersymmetric Yang--Mills theory is \cite{B2,OT,A3}
\begin{gather}
\begin{aligned}
b=&-\frac{1}{2}\xi^{-1}(\lb\g^aD)\partial_a
+\frac{1}{16}\xi^{-2}(\lb\g_{abc}r)\Big(N^{ab}\partial^c-\frac{1}{24}D\g^{abc}D\Big)-\\
&-\frac{1}{64}\xi^{-3}(r\g_{abc}r)(\lb\g^aD)N^{bc}-\frac{1}{1024}\xi^{-4}(\lb\g_{abc}r)(r\g^{cde}r)N^{ab}N_{de}
\end{aligned}\label{eq.bSYM}
\end{gather}
where it is useful to note that the last term in $b_1$ is proportional to $(r\g^aD)(\lb\g_aD)$. As $(\lb\g^a)^\alpha(\lb\g_a)^\beta=0$ and $b_2$ encodes $\lb_{[\alpha} r_\beta r_{\gamma]}$ we then have that any pair of $\{D_\alpha,D_\beta\}$ constructed out of these terms gives a zero result. Similarly, none of the $N^{(2)}$ acts on $\xi$, so it is no surprise that $bb=0$. In addition, the original $\partial$:s show up as $(\lb\g^a)^\alpha\partial_a$ and therefore cannot form $\partial^2$.

For the formation of $\partial^2$ it is essential that $r_\alpha$ provides\footnote{In this way $r\sim D\sim\partial^{1/2}$ and $b\sim r^3$, while the corresponding for the propagator is $r^{-1}$.} $(\la\g_{ab}D)(\lb\g^{ab})_\alpha$ by the loop regularisation. At the introduction of the regularised $D$:s, new $\partial$:s can be formed, in effect out of $r^2$ and $rD$. Moreover, since the former shows up in the configuration of $(\la\g^{a})_\alpha\partial_a$ and the latter in $(\lb\g^a)^\alpha\partial_a$, a formation of $\partial^2$ in the regularised setting demands the equivalence of $r^3D$ or $r^2\partial$. Therefore, the restriction\footnote{By \eqref{eq.comps}, $[\mathrm{d}D^{I}]$ and the circumstances of $b_j$.} of the $D$:s and $\partial$:s available to a number of $L-6+j$ for $L\geq2$ and $j-4$ for $L=1$ constrain the $\partial^2$:s possibly present to the same amount. Consequently, given \eqref{eq.UVfinC}, we can immediately conclude that supersymmetric Yang--Mills theory is finite in the ultraviolet regime for dimensions
\begin{equation}\label{eq.DSYM}\left.\begin{array}{ll}
D<8,&L=1\\
D<4+\frac{6}{L}, & L\geq2
\end{array}\right.\end{equation}
in consistency with what is well known \cite{Green:1982sw,Howe:2002ui,Bern:1997nh,Bern1,Bern:2010tq,Bern:2012uc,Bern:2012di}. The analysis may also be performed with respect to $r$, giving further restrictions for large $j$.

\section{The nature of the loop regularisation}\label{par.loopreg}
The introduction of a regulator with exponent as in \eqref{eq.loopreg} together with an integration over the loop momenta ensures enough non-locality of the propagator to allow for the construction of examinable loop structures. It is not without complications though. The most obvious issue is the conversion of $r$ into $\la\lb D$. By $Q$-equivalence, any regularised expression is equivalent to the initial one. However, combinations of unregularised terms often result in expressions proportional to zero, while divergent with respect to small $(\la,\lb)$, including singularities concerning $\eta$ in supergravity. As $(0/0)$ makes no sense we need to consider the regularised operators to capture the full theory, thereby introducing extra $(\la,\lb)$ and finite results with respect to small $(\la,\lb)$.

Clearly, sufficient regularisation in this context (as well as for tree diagram parts, under general regularisation) provides finite terms\footnote{Appendix \ref{app.loopfinite} describes the extent of this for loop clusters in Yang--Mills theory.} with respect to small $(\la,\lb)$. Further conversions of $r$ are not necessary for a full analysis, where vanishing results are allowed for. Moreover, the highest surviving power of $r$ in a regularised expression by $Q$-equivalence encodes the extra terms originating in the regularisation of that $r^x$ (not contingent upon the survival of the lower terms).

So far, e.g. in \cite{BG,JB,CK2}, this has been interpreted as though it were only necessary to consider the regularisation of $r$ when the number present exceeded its degrees of freedom (11 in Yang--Mills theory, 23 in supergravity). It has been assumed to constitute the point at and below which terms do not (with loop regularisation: wrongly) vanish, that is: due to too high a power of $r$.

This now seems too naive, the crucial observation of which is that the operators in the loop clusters in Yang--Mills theory and supergravity cannot form $(r^{11},r^{23})$. The former is the case most easily illustrated. Out of the configurations of $(\lb,r)$ in \eqref{eq.bSYM} at most $r^{10}\lb^{n}$ ($n>1$) can be formed\footnote{For any analysis concerning $r^x\lb^y$, their irreducible representations are very useful. These are listed in \cite{CK1}, for the case of maximally supersymmetric Yang--Mills theory, together with a LiE program for how to compute them. Note that the building blocks, with $\lb$ in (00010), are the fermionic $\lb r$ in (00100), the bosonic $\lb r^2$ in (01001) and the fermionic $\lb r^3$ in (02000). The spinors in each set are antisymmetrised, and the $[\mathrm{d}r]$ in \eqref{eq.dr} require $r^3$ in (00030), replaceable by $\lb^3$.}. The components of the $b$-ghost simply cannot support a configuration with $r^{11}$. Since the only other $r$:s present occur in external states and at the final superspace integration, and therefore do not show in our analysis of the loop clusters, this means that $r^{11}$ is absent for our purposes.

The corresponding for maximal supergravity is slightly more involved, but it is possible to note that the $(\lb,r)$ in the operators \eqref{eq.oSUGRA} combine in a specific way \eqref{eq.L+} as to only be compatible with the final $[\mathrm{d}r]$ for up to $r^{15}$. Any higher power of $r$ out of the configurations provided by the operators concerned cannot be combined with antisymmetrised $r$:s to form $r^{23}$ in the irreducible representation of $(02003)$ which is required by \eqref{eq.dr} \cite{Cederwall:2009ez}. This occurs due to the set combinations of $(\lb,r)$ of the operators and the subsequent $\lb^nr^{23}$ of the final expression, where it must be possible to replace $r^{23}$ with $\lb^7$, as specified by \eqref{eq.dr}, with a non-zero result. The relevant $\lb^nr^{15}$ can be found in \eqref{eq.L(15)}.

The absence and divergence\footnote{Each $r$ is accompanied by $\xi^{-1}$ in Yang--Mills theory, and by $\sigma^{-2}$ in supergravity.} (in the loop clusters) of
\begin{equation}\label{eq.rL1015}
(r^{x>10},r^{x>15})
\end{equation}
calls for a regularisation of $r$ down to a power of $(10,15)$, constituting non-zero configurations of $(\lb,r)$. This is sufficient, as any $f(\la,\lb,r)$ can be regularised (loop/general) to finiteness with respect to small $(\la,\lb)$. The subsequent highest non-zero configuration $\la^x\lb^yr^z$ (with loop regularisation: the highest $z$) encodes the further regularisations\footnote{In line with this, we may observe that none of the components of a regularised $r$ is required by loop integration, as noted for Yang--Mills theory in \S\ref{par.lUV} and for supergravity in \S\ref{par.csD}, i.e. the non-zero configuration of $r^x\lb^n$ may equivalently remain.}. Divergent terms in this way remaining after loop integration (of an entire loop cluster) are taken care of by the general regularisation.

We will now continue by observing some key characteristics of the amplitude diagrams, due to the loop regularisation and the integration over the loop momenta. The main points concern $Q$-equivalence, total derivatives and loop momenta claimed by the loop integrations.

\subsection{Characteristics of the loop regularisation}\label{par.charL}
Important to note is that the regulators present do not alter basic properties such as\footnote{There is unfortunately an incorrect statement on this in \cite{CK2}.} $bb=0$. Regardless of how many terms each $b$ has been converted into, once two full sets (including terms that may give a zero result at the loop integration) show up next to each other (on the same propagator), the result is zero. In general, if no parts of two operators $A$ and $B$ have acted out on other propagators in the process of ending up next to each other, we have
\begin{equation}
e^{\{Q,\chi\}}Ae^{-\{Q,\chi\}}e^{\{Q,\chi\}}Be^{-\{Q,\chi\}}=e^{\{Q,\chi\}}ABe^{-\{Q,\chi\}}.
\end{equation}
if so is allowed by $Q$-equivalence, which is described below.

A second important observation has to do with general loop properties concerning derivatives: a derivative in a loop remains there, without acting on anything, if so is possible. Otherwise, the derivative in some way acts on another component: derivative, variable or across a vertex, i.e. out of the loop.

\subsubsection*{Total derivatives}
The consequences of the two observations above are more far-reaching than perhaps assumed at a fist glance. The fact that $bb=0$ prohibits two full $b$-ghost to remain in a loop has implications in combination with the loop integrations, which claim derivatives and make use of the loop regulator, bringing down other derivatives and variables. Each loop integration claims $\omega$, $\bar{\omega}$ and $s$ to a number of their degrees of freedom $(11,23)$, whereas $D$:s are claimed to a full set of fermions $(16,32)$. The integrations over $\partial$ claim none, but result in the ultraviolet divergences described in \eqref{eq.UVfinC}.

The required $\bar{\omega}$:s are provided by the loop regulator \eqref{eq.loopreg} as no other operator contains any, compare \eqref{eq.bSYM} and \eqref{eq.oSUGRA}. This also brings down a full set of $\omega$ on one inner propagator of the loop in question, a propagator which may be considered to be the gathering point for the loop components at the loop integration. Moreover, the other $\omega$:s present in the loop are not forced to act on anything (i.e. remains in the loop as a derivative) by any other relations than $bb=0$ and (possibly) $(\la\g_{ab}\la)[R^a,R^b]$. Any $\omega$ not forced to act on something, possibly out, is added to the $\omega$:s brought down by the loop integration, and represents a total derivative at the loop integration: a vanishing expression. In addition, since $bb=0$ only forces one derivative in $b$ out of a loop (as that $b$ acts across a vertex), splitting the $b$, only $b_j$:s  contain $\omega$ (maximum one) with a non-zero result. The $R$:s on the other hand effectively act into different loops (with respect to $\omega$). For the outer vertices, one $R$ can be equivalently considered to act out of the loop cluster, simplifying the analysis. Only these $R$:s carry $\omega$ with a non-zero result, and since they act out of the loop cluster, they do not affect the UV divergences. In total, this diminishes the number of operator components necessary to consider as parts of the loop cluster.

The requirements of $s$ and $D$ are less dramatic. A loop integration over $s$ claims the required loop momenta from the loop regulator, for the same reasons as for $\bar{\omega}$, which subsequently brings about the same number of $D$:s $(11,23)$. In Yang--Mills theory, these must be combined with $D^5$ from the operators (original, not brought down by the loop regularisation) for the loop integration over $D$ to be satisfied. No other combination is accepted, and other $D$:s in the loop are forced to act on other components. For supergravity, at least $9$ original $D$:s are required, though the end composition might possibly vary between $9$ and $11$, which is further commented on in \S\ref{par.csD}.

An obvious, possible result of $D$ being forced to act on other components, like another $D$, is $\partial$. The corresponding loop integration does not claim any loop momenta, but result in UV divergences as described in connection to \eqref{eq.UVfinC}. Important to note is that unpaired $\partial_a$:s remain in a loop, if not part of $b_j$ and forced out of the loop cluster akin to $\omega$. In the loop cluster, they give a vanishing result by the combination of representing either a total derivative (when not constituting a momenta of that loop) or an odd function. Only $\partial^2$:s survive in the loop clusters, and from the next paragraph and \eqref{eq.oSUGRA} we can conclude that any term containing a $\partial_a$ that cannot combine into $\partial^2$ may be disregarded.

\subsubsection*{$Q$-equivalence}
In the analysis of a part of a diagram, $Q$-equivalence allows for up to $(r^{10},r^{15})$ before loop regularisation needs to be considered. Moreover, any such examined entity can be regularised, and the observed properties of the initial parts carry over to the regularised combination, equivalently combined after regularisation.

For example the examination of two $b$-ghosts on each side of a 3-point vertex can be carried through in this manner (compare \eqref{eq.bSYM} and \eqref{eq.oSUGRA}). The process of one of the $b$:s acting across the vertex and the result in relation to the other $b$ therefore can be analysed without prior loop regularisation, though the $r$:s better not be considered to leave their original propagators. If the two $b$:s make up part of a loop, the property of $bb=0$ then forces one derivative of each term in the acting $b$ onto the third leg. Moreover, a regularisation of the split $b$ is $Q$-equivalent to the expression of a regularised $b$, split due to $bb=0$. So it comes to be that any $b_j$ at most contributes with one original $D$ or $\partial$ to the loop structure; compare with the relevant terms of $b_0$ and $b_1$ in \eqref{eq.bSYM} and \eqref{eq.oSUGRA}.

\section{Maximal supergravity}\label{par.msugra}
In $D=11$, $\mathcal{N}=1$ supergravity the spinors are symplectic, making it convenient to use an implicit $\varepsilon^{\alpha\beta}=\varepsilon^{[\alpha\beta]}$ to capture the way spinor indices are connected to each other\footnote{This affects ordering. The spinors have one chirality, with ($r_\alpha,\lb_\beta$) e.g. as $(\lb r)=-(r\lb)$.}. In order to further express the operators in a simple form, we use
\begin{equation}\label{eq.L}
L^{(p)}_{a_0b_0,a_1b_1,\ldots,a_pb_p}=(\lb\g_{\lbracket a_0b_0}\lb)
(\lb\g_{a_1b_1}r)\ldots(\lb\g_{a_pb_p\rbracket}r),
\end{equation}
where $\lbracket\ldots\rbracket$ denotes antisymmetrisation between the $p+1$ pairs of indices. This tensor has the properties of
\begin{align}
L^{(n)}L^{(m)}&\propto(\lb\g^{(2)}\lb)L^{(n+m)}\label{eq.L+},\\
[\bar{\partial},\eta^{-(p+1)}L^{(p)}_{a_0b_0,\ldots,a_pb_p}\}&=2(p+2)\eta^{-(p+2)}L^{(p+1)}_{ab,a_0b_0,\ldots,a_pb_p}(\la\g^{ab}\la),
\end{align}
where $\bar{\partial}=r\bar{\omega}$. The operators, i.e. the $b$-ghost\footnote{We here present a different $b$-ghost compared to that of \cite{CK2} in order to facilitate the analysis of the theory. By fulfilling \eqref{eq.bReq} it provides a $Q$-equivalent alternative. However, note that the b-ghost in \cite{CK2} suffered from sign errors; the overall signs for the $b_2$ and $b_3$ should have been negative.}
and the operators in the vertices: $R^a$ and $T$, are then possible to express as:
\begin{subequations}
\begin{gather}\begin{aligned}
b=&\frac{1}{2}\eta^{-1}(\bar{\la}\g_{ab}\bar{\la})(\la\g^{ab}\g^iD)\partial_i+\\
&+\eta^{-2}L^{(1)}_{ab,cd}\Big((\la \g^{a}D)(\la \g^{bcd} D) +2(\la{\g^{abc}}_{ij}\la)N^{di}\partial^j+\\
&\qquad\qquad+\frac{2}{3}(\eta^b_p\eta^d_q-\eta^{bd}\eta_{pq})(\la\g^{apcij}\la)N_{ij}\partial^q\Big)-\\
&-\frac{1}{3}\eta^{-3}L^{(2)}_{ab,cd,ef}\Big((\la\g^{abcij}\la)(\la\g^{def}D)N_{ij}-\\
&\qquad\qquad-12\Big[(\la{\g^{abcei}}\la)\eta^{fj}-\frac{2}{3}\eta^{f[a}(\la\g^{bce]ij}\la)\Big](\la\g^{d}D)N_{ij}\Big)+
\end{aligned}\end{gather}
\begin{equation}
+\frac{4}{3}\eta^{-4}L^{(3)}_{ab,cd,ef,gh}(\la\g^{abcij}\la)\Big[(\la{\g^{defgk}}\la)\eta^{hl}-\frac{2}{3}\eta^{h[d}(\la{\g^{efg]kl}}\la)\Big]\{N_{ij},N_{kl}\}\nonumber
\end{equation}
\vspace{-0.4cm}
\begin{gather}
\begin{aligned}
R^a=&\eta^{-1}(\lb\g^{ab}\lb)\partial_b-\eta^{-2}L_{(1)}^{ab,cd}(\la\g_{bcd}D)+\hspace{4.1cm}\\
&+2\eta^{-3}L_{(2)}^{ab,cd,ef}\Big[(\la\g_{bcdei}\la)\eta_{fj}-\frac{2}{3}\eta_{f[b}(\la\g_{cde]ij}\la)\Big]N^{ij}
\end{aligned}
\end{gather}
\begin{equation}\label{eq.T}
T=8\eta^{-3}(\lb\g^{ab}\lb)(\lb r)(rr)N_{ab}\hspace{6.0cm}
\end{equation}\label{eq.oSUGRA}
\end{subequations}
Here, we may immediately note that $T$ ``has to'' act out of the loop cluster. Otherwise the $\omega$ would constitute a total derivative, with a vanishing result. This limits 4-point vertices to be present in loop clusters as structures with the one property of providing extra outer legs. A $j$-point loop cluster made out of 3-point vertices can be changed into a $(j+1)$-point loop cluster through the addition of one outer leg to a vertex, a process which changes none of the properties of the loop cluster, apart from eventual $(\lb r)^2=0$, as the additional term is outside of it. No extra conditions on the operators inside are introduced. Because of this, it is sufficient to examine loop clusters made solely out of 3-point vertices for a complete picture of the UV divergences of the amplitudes, and we will therefore not deal with 4-point vertices, or $T$, any further in this article.

Secondly, as the operators in the vertices act on any leg in an equivalent manner (as long as the legs are separate), each outer vertex can be interpreted as providing one $R^a$ acting out of the loop cluster, and the other one in. The former does not contribute to the divergences of the loop cluster, and the components of the latter that give vanishing contributions may be disregarded, same as goes for all operators of the inner vertices. For example, the $R_2$ is ``absent'' due to its containing $\omega$. The $R_0$ on the other hand contains  $(\lb\g^{ab}\lb)\partial_b$, which cannot combine into $\partial^2$, as we soon will show.

\subsection{The effective operator contributions inside a loop}
We will now examine the terms of $b$ and $R^a$ for further observations concerning what does (not) contribute to the ultraviolet divergences, with respect to the formation of $\partial^2$ and the combination of $D$ into $D^{32}$ per loop, to satisfy $[\mathrm{d}D^{I}]$. To begin with the former, we may note that any pair of original $D$:s from the operators above, acting on each other, vanishes\footnote{For the relations involved, see appendix \ref{app.id}, in particular the last equation of \eqref{eq.Leq}. It gives at hand that the antisymmetrisation of the $L^{(n)}$-indices is encoded by the irreducible representations of the rest of the expression, with the indices contracted with $\lb\g^{mn}\lb$ in a position of choice. For the purpose of original $\{D,D\}\propto\partial$ we have the three first terms in \eqref{eq.listofD}.}. Including regularised contributions, we have a set of $D$ up for combinations:
\begin{equation}\label{eq.listofD}
(\la\g^{mn}\g^iD)\partial_i\quad(\la\g^m D)\quad(\la\g^{imn}D) \quad(\la\g^{jm}D)\quad (\la\g^{mn}D),
\end{equation}
indices contracted with $\lb\g^{mn}\lb$ and where a contraction of the $j$ with a corresponding $m$-index yields a vanishing result. The two last terms correspond to regularised $r$:s, by \eqref{eq.rTransf}:
\begin{equation}\begin{array}{rl}
(\lb r): &\text{ }\text{ }(\lb\g_{mn}\lb)(\la\g^{mn}D)\\
(\lb\g_{\lbracket ab}\lb)(\lb\g_{cd\rbracket}r): &4(\lb\g_{\lbracket ab}\lb)(\lb{\g_{c}}^m\lb)(\la\g_{d\rbracket m} D)
\end{array}\end{equation}
These do not act on the $D$:s or converted $r$:s in the $b_n$ or $R^a_n$ they originate from, nor on the configuration of operators originally on their propagator (at most $RbR$), as the original expression effectively contains $[r,D]$ etc.\footnote{The relevant terms are possible to analyse in a $Q$-equivalent manner, without regularisation.}

The first term in \eqref{eq.listofD} acting on $D_\alpha$ gives $\propto\partial^2(\la\g^{mn})_\alpha$ which cannot be combined with any of the terms above. After that, it is suitable to observe that any $\partial_m$ cannot form $\partial^2$ with a $\partial$ formed out of the $D$:s above, as this would result in $[mnn']$. Since $\partial_m$ cannot combine with a $\partial_n$ or any of the original $\partial$:s either, we have that it gives a vanishing contribution, and it may be disregarded in further examinations\footnote{$\partial_m$ may only exist in $b_0$ as forced to act out of the loop cluster.}. Then the fifth term cannot act on any other $D$ either, and only the latter part of the third: $(\la\g^{imn}D)=(\la\g^i\g^{mn}D)+\eta^{im}(\la\g^{n}D)$, which comes with the formation of $\partial_m$, $(\lb r)$ or its regularised counterpart (the fifth term in \eqref{eq.listofD}). Nor can two terms of the second kind form $\partial$.

The vanishing of $\partial_m$ and $\omega$ except on $b_j$ gives at hand that we in loops, regarding the examination of non-zero contributions, effectively have the operators\footnote{Compare with \eqref{eq.oSUGRAw}.}:
\begin{subequations}
\vspace{-0.5cm}
\begin{align}
b&^{\text{eff.}}_{\text{loop}}=\frac{1}{2}\eta^{-1}(\bar{\la}\g_{ab}\bar{\la})(\la\g^{iab}D)\partial_i+\nonumber\\
&+\eta^{-2}L_{ab,cd}^{(1)}\big[(\la\g^aD)(\la\g^{bcd}D)+8(\la\g^{a[s}\la)(\la\g^{bcd]}\omega)\partial_s\big]+\\
&+\eta^{-3}L^{(2)}_{ab,cd,ef}\big[6(\la\g^{def}D)(\la\g^{[ab}\la)(\la\g^{c]}\omega)-16(\la\g^{f}D)(\la\g^{a[b}\la)(\la\g^{ecd]}\omega)\big]\nonumber\\
(&R^a)^{\text{eff.}}_{\text{loop}}=-\eta^{-2}L_{(1)}^{ab,cd}(\la\g_{bcd}D)
\end{align}
\end{subequations}
though it is important to remember that there are implicit terms (with vanishing contributions) in the description, taking care of e.g. $bb=0$.

We may also observe that the only non-zero $\partial$-contributions come from $b_0$, $b_1$, $rr\rightarrow\partial$ and $r(\la\g^{m}D)\rightarrow\la\partial$. However, $\partial^2$ seems possible to form out of any combination of these structures, such as $\partial^2$ (unregularised) and $r^4$:
\begin{subequations}\begin{align}
&r^2\rightarrow\partial:\quad\{(\la\g^{im}D),(\la\g^{jn}D)\}\propto(\la\g^{ijmns}\la)\partial_s\label{eq.2r2p}\\
&r^4\rightarrow\partial^2:\quad(\la{\g^{ijmn}}_s\la)(\la\g^{kl\bar{m}\bar{n}s}\la)\partial^2\propto(\la\g^{\bar{m}\bar{n}}\la)(\la\g^{kl}\g^{ijmn}\la)\partial^2
\end{align}\end{subequations}
This complicates the deduction of the restrictions on $\partial^2$ that ought to occur, though an important point is that the units of $(\la\g^{imn}D)(\lb\g_{it}r)$ at most can contribute the worth of one $D$ (original or regularised) to the formation of $\partial^2$, and in effect $\partial$ itself.

\subsection{The consequences of the loop integrations over $s$ and $D$}\label{par.csD}
At the integration over $s$, a total of  23 $(\la\g_{ab}D)(\lb\g^{ab}s)$ and $(\la D)(\lb s)$ (at most one of the latter) are provided by the loop regulator, as $s^{23}$ in the irreducible representation of (02003) needs to be claimed by $[\mathrm{d}s]$ for a non-zero result. For each loop, these can be interpreted to show up on one inner propagator. There, the momenta effectively belong to that loop only, and it is where that loop's momenta effectively can be considered to congregate at the loop integration, as the loop turns into a vertex. It is equivalent to assume a set of $RbR$ to exist on that propagator (no components acting on one another) prior to $(\la D\lb s)^{23}$ being singled out. Moreover, the (part of the) regulator providing these shows as e.g. $\{e^{\{Q,\chi\}},r\}$ ($r$ already considered to be regularised) and effectively:
\begin{equation}
\{(\la D\lb s)^{23},e^{\{Q,\chi\}}re^{-\{Q,\chi\}}\}\propto[(\la D)^{23},e^{\{Q,\chi\}}re^{-\{Q,\chi\}}](\lb s)^{23}.
\end{equation}
Consequently, the $(\la D)^{23}$ brought into the loop by the integration over $s$ does not act on any $D$ initially on the propagator, originating in a regularisation or not. Those $D$:s must be antisymmetrised with the $D^{23}$, and each other.

Now, $(\la D)^{24}$ is the maximal antisymmetrisation of 2- and 0-form $\la D$:s that can be formed, consisting of 24 $\la\g^{(2)}D$ or 23 $\la\g^{(2)}D$ and one $(\la D)$, in the irreducible representations of $1\mathrm{x}(05006)+1\mathrm{x}(06004)+1\mathrm{x}(07002)+1\mathrm{x}(08000)$. However, for the integration over $D$ these in combination with $\la\g^{m}D$ (at most 2), $\la\g^{imn}D$ and the $\la^2D^2$ of $b_1$ must form $\la^{23}D^{32}$ with irreducible representations of $\la^{32}$. This is not possible. For the integration over $D$ to provide a non-zero expression, at most 23 2- and 0-form $\la D$ can contribute to the $D^{32}$.

This means that any $r$ initially on the propagator can be regularised, thus providing finiteness with respect to small $(\la,\lb)$. However, only the original $r$ in the regularised expression gives a non-zero contribution at the loop integration. Moreover, for each loop we have at least $r^2$ from the $R$:s on the propagator (sometimes $r^3$, if $b_1$ is required), so a minimum of $r$:s present in a loo cluster is always $2L$, a number which cannot exceed 15 with a non-zero result. Consequently,
\begin{equation}\label{eq.L7}
L\leq7 
\end{equation}
for loop clusters\footnote{This might seem odd, but the loop regularisation gives at hand an expression regularised with respect to small $(\la,\lb)$. If too many loops are present in the loop cluster, the loop integrations then set the expression to zero.}. Amplitude diagrams containing more loops than that do so with the loops divided upon several loop clusters. Interestingly, this gives that an amplitude diagram does not behave worse than at seven loops.

A second consequence of the maximal contribution of 2- and 0-form $\la D$ to the integration over $D$ is that at least 9 original $D$:s are claimed by each loop integration. As mentioned, and obvious from $[mnn']=0$, at most two of these\footnote{The corresponding for $\la\g^{imn}D$ is $9$, but then of course there is also the combination of $\la^2D^2$ in $b_1$, where the indices $mn$ can be moved at will by \eqref{eq.Leq}.} can come in the shape of $\la\g^m D$. For example, the absence of bubbles and triangles follows directly from this, even without the consideration of $Q$-equivalence, since a loop with $j$ propagators thus cannot contribute with more than $2j+2$ original $D$:s to the loop integration. A closer examination including $Q$-equivalence, as in the last part of \S\ref{par.charL}, sets the maximal number of original $D$:s provided by the loop components to
\begin{equation}\label{eq.maxOD}
\begin{array}{ll}
1+2j,&L=1\\
10+2j,&L=2\\
9L-7+2j,&L\geq3,
\end{array}
\end{equation}
since at most one original $D$ is provided by each $b_j$. In that way, at most $D^2$ is provided per vertex (possibly both $\la\g^{imn}D$) while the inner propagators through $b_1$ may yield $(\la\g^{m} D)(\la\g^{nab}D)$ until the fist part hits its maximum of $2L$. 

The loop integration requirement of $D^{9L}$ (original, covariant derivatives) in combination with \eqref{eq.maxOD} gives the requirement of
\begin{equation}
j\geq4
\end{equation}
for \emph{any} loop cluster in maximal supergravity, not merely for single loops as by \eqref{eq.jLimGen} in Yang--Mills theory. The presence of $b_1$ has further consequences though. For the 4-point, $L\geq 3$ loop cluster, these must amount to at least $2L-1$ on the inner propagators for the $[\mathrm{d}D^{I}]$:s to be satisfied. This gives a maximum of $L-2$ inner propagators free of $b_1$, resulting in a minimum of two cases of $r^3$ on the inner propagators subject to $[\mathrm{d}s^{I}]$ and $(\la\lb D s)^{23}$ from the loop regulator\footnote{The other side of the coin is that with $x$ $b_1$:s on these inner propagators, the other $L-x$ ones must constitute $b_0\propto\partial$, each forming $\partial^2$ within their own loop integration.}. The requirement for a non-zero result, modified compared to above, follows as
\begin{equation}\label{eq.L6}
j=4 \quad\Rightarrow\quad 2L+2\leq15 \quad\Leftrightarrow \quad L\leq 6.
\end{equation}
The 4-point loop cluster is only supported up to 6 loops; the 7-loop cluster requires at least five outer legs.

\subsection{The ultraviolet divergences}\label{par.UVdiv}
A complication with respect to the examinations of the ultraviolet divergences in maximal supergravity is that more components can form $\partial^2$ than what is true for the case of Yang--Mills theory. The entity can be constructed out of $r^4$ as well as original $\partial$:s. Moreover, the observed limits on what is allowed in connection to the loop integrations are few, and it seems like there should be more to discern from the formalism in that respect.

Out of the structures of the operators, the most important features to be noted in respect to the limits on what may form $\partial^2$ are the vanishing of $(\lb\g^{mn}\lb)\partial_m$ and the property that  $(\lb\g^{mn}\lb)(\la\g^{imn}D)(\lb\g_{it}r)$ only can provide the worth of one $D$ to $\partial^2$:s. The last one inevitably means that for large $j$, the addition of an outer leg does not limit the requirements on the dimension (for finiteness) any further, since it at most adds the equivalent of $D^4/\partial^2$ to the loop components that may form $\partial^2$. However, for low $j$ it provides no useful information, even if both $r$:s in $L^{(2)}$ might be affected. It merely infers, in combination with that $b_0$ at most can provide $D^2$ (by $\partial$) to the loop structure, that the worth of $D$ from the inner propagators and that of $D^2$ from the vertices (using $r\sim D\sim \partial^{-1/2}$) are prohibited from affecting the divergences. However, these $7(L-1)+j$ $D$:s are already known to be removed from the loop cluster\footnote{As is the $r^{2L}$, or e.g. $r^{2L+2}$ for $L>1$, $j=4$, that remains after loop regularisation.} by the $[\mathrm{d}D^{I}]$:s $(9L)$ and the recognised presence of up to $r^{15}$, before the conversion of $r$ into $D$ by loop regularisation needs to be considered.
 
A definite difference in comparison to Yang--Mills theory is that $\partial^2$ might be formed in loop clusters not subject to the conversion in \eqref{eq.rTransf}, as the original $\partial$:s seem to be able to combine. However, these occur to a limited extent due to the requirement of original $D$:s described in \eqref{eq.maxOD} in combination with that $b_j$ at most contributes with one original $D$ or $\partial$. For the $4$-point amplitude they are limited to exist for $L>3$, on $L-2$ inner propagators (the only ones not occupied by $b_1$) or on $L-3$ inner and one $b_j$. The $\partial^2$:s formed in this way are limited to\footnote{$[x]$ representing a rounding of $x$ to the closest lower integer.}
\begin{equation}\label{eq.pUnreg}
[(L-2)/2],
\end{equation}
if even that, considering that these cases often would constitute a $\partial$ remaining in the loop. Additional outer legs do not worsen the ultraviolet behaviour set by \eqref{eq.pUnreg}, since their presence at most adds and frees $\partial$ (with respect the $[\mathrm{d}D^{I}]$) enough to compensate for the $1/\partial^2$ accompanying each propagator. In total, the possible formation of one $\partial^2$  for $4\leq L\leq5$ and two for $6\leq L\leq7$ even fails to compensate for the requirement of additional outer legs in supergravity, compared to Yang--Mills theory. The property of $\partial^2$ forming from original $\partial$:s as such causes no worse an ultraviolet divergence than that of unregularised Yang--Mills theory. The relevant examination concerns loop clusters with regularised $r$:s and then, the above is included per default.

A worst case scenario with as many parts of the regularised operators as possible acting into $\partial^2$ would look like \eqref{eq.UVfinC}
\begin{equation}
LD-6(L-1)-2j+2\bigg[\frac{9(L-1)+8(L-1)+5j-9L-x}{4}\bigg],
\end{equation}
$x$ denoting the components not party of the formation of $\partial^2$ after the removal of $D^{9L}$ for the loop integration, and the expression within brackets being required to be positive. At present, our highest confirmed $x$ is that of $15$, the default for regularised loop components. The subsequent limit on the dimension fails to give a conclusive result. By it, a 4-point amplitude is finite in the ultraviolet regime if
\begin{equation}\begin{array}{ll}
D<8,& L=1\\
D<2+\frac{8}{L},&L\geq2.
\end{array}\end{equation}
Compared to the known limits on the dimension for finiteness in maximal supergravity, coinciding with maximally supersymmetric Yang--Mills theory up to $L=4$ etc., what this tells us is that the are more restrictions on the terms combining into $\partial^2$ than observed so far.

It is either pure coincidence or a curious fact that the removal of $r$ to its degrees of freedom, 23, comes very close to the ``anticipated'' result of \eqref{eq.DSYM} up to $L=4$ \cite{Bern2,Bern25,Bern:2012uf} and $D<2+14/L$ \cite{BG,JB} above it, in agreement with \cite{BHS,Vanhove:2010nf} (it differs slightly by the contributions caused by $j\geq4$). The recognised, effective extra degrees of freedom of $r$ (between $r^{16}$ and $r^{23}$), by the loop regularisation, after all is not caused by $r$ exceeding its degrees of freedom, but is due to the configurations of the variables (irreducible representations etc.). It would be interesting to know if this affects the contribution of converted $r$:s to the loop integrations.

In that case, there is likely an additional constraint concerning the contributions from $j$ too. As may be noted, the parts of $b_j$ not proportional to $\omega$ cannot contribute with more than the worth of $D^2$ to the loop cluster, which currently is not the case for the other parts.

\section{Conclusions \& outlook}
In the pure spinor superfield formalism, a key feature in determining the actual ultraviolet divergences of amplitude diagrams is how far terms in one-particle irreducible loop structures (loop clusters) need to be regularised with respect to small $(\la,\lb)$. A regularisation to finiteness gives the complete picture, allowing for zero results, but it complicates the analysis in an unnecessary way, sometimes requiring a general regularisation in addition to the loop regularisation. Configurations of $(\la,\lb,r)$ encode further regularisations by $Q$-equivalence, provided they are non-zero.

During loop regularisation, the key component of $(\la,\lb,r)$ is $r$ and the highest power of it that can be present. This has been assumed to be constituted by its degrees of freedom: 11 in maximally supersymmetric Yang--Mills theory and 23 in maximal supergravity. However, an examination of what can be provided by the operators and regulators of the loop structures shows that $r$ cannot be provided to a higher power than 10 in Yang--Mills theory, and 15 in supergravity.

To what extent this affects previous studies using pure spinors in maximally supersymmetric Yang-Mills theory like \cite{BG,JB} is difficult to tell, but as the change of the limit is slight, the consequences might be negligible. However, for maximal supergravity, the assumption of $r^{23}$ \cite{CK2} is completely wrong, unless the regularisations past the degrees of freedom of $r$ somehow do not affect the ultraviolet divergences.

The presumption of $r^{15}$ to set the limit of what is necessary to consider in relation to the ultraviolet divergences of the loop clusters is not enough to give results as predictive as those of other studies, with respect to what dimensions the theory is finite in. For results such as the 4-pt requirements of $D<8$ for $L=1$, $D<4+6/L$ for $2\leq L\leq4$ \cite{Bern2,Bern25,Bern:2012uf} and the limit of  $D<2+14/L$ for $L\geq5$ \cite{BG,JB}, more examinations of the formalism are required.

Most importantly though, the pure spinor formalism states a cut-off of the loop dependence at $L=7$, due to the loop integrations and the maximal power of $r$ as 15. That is, loop clusters of more than seven loops cannot be formed, and those with seven loops demand five outer vertices to be connected to them (four being the default requirement). Diagrams with more than seven loops must be one-particle reducible for a non-zero result.

This is especially interesting in combination with the previous predictions for $L\geq5$ of \cite{Vanhove:2010nf,Beisert:2010jx}. The combined results imply a finiteness of 4-point amplitudes in $D=4$, possibly with a logarithmic divergence for amplitudes with more external states, especially if the results of \cite{BG,JB} hold. At that point, it becomes important if another property of maximally supersymmetric Yang--Mills theory carries over to maximal supergravity: the softening of the limit on the dimension at an increase of the number of states directly attached to the loop cluster. That is crucial for determining the ultraviolet divergences in $D=4$, i.e. if there is a logarithmic divergence at seven loops.

It now seems quite likely that maximal supergravity in $D=4$ at most diverges logarithmically. For a confirmation of this in a pure spinor setting, a study of the actual combination of terms in the loop clusters is required, in addition to the conceptual analysis above. It is not obvious that merely the presence of manifest supersymmetry is enough to yield the correct results; perhaps manifest U-duality is required as well. In general, constraints are difficult to discern, though the loop integrations over $D$ possibly provide restrictions on the UV divergences. The contributions from the outer legs are also critical for general predictions; in fact, that is likely to be true for the 4-point diagrams as well.

\section*{Ackowledgements}
I would like to thank M. Cederwall for helpful discussions.

\appendix
\section{Finite loop clusters in Yang--Mills theory}\label{app.loopfinite}
Here, we will have a quick look at how far a loop cluster needs to be regularised for finiteness with respect to small $(\la,\lb)$ in maximally supersymmetric Yang--Mills theory, in principle following the argumentation of \cite{JB}. The corresponding for supergravity remains to be performed.

In short, the contribution is $\lb/\xi$ from the propagators, and $r/\xi$ from the $r$:s, which is possible to deduce from \eqref{eq.bSYM}. A further identifiable characteristic is that each loop integration gives at hand $\la^{3}$ (each $[\mathrm{d}N]$ claims $8$, but each $[\mathrm{d}S]$ brings down $11$ from the loop regularisation). In the presence of $3(L-1)+j$ propagators the end result is:
\begin{equation}\label{eq.xiSYM}
\la^3\bigg(\frac{\lb}{\xi}\bigg)^j\bigg(\frac{r}{\xi}\bigg)^x
\end{equation}
Of these, the first $\la^3$ needs to remain and be claimed by $[\mathrm{d}\la]$ \cite{B2}, and a regularisation of $r$ as in \eqref{eq.rTransf} takes care of the divergence coupled each such variable. In conclusion, in maximally supersymmetric Yang--Mills theory a loop regularisation 
\begin{equation}\label{eq.rnumberSYM2D}
r^y\rightarrow (\lb\la D)^y:\quad y=x+j-10,\quad x+j\leq10
\end{equation}
is necessary for the loop cluster to provide an finite entity. As can be observed, this does not suffice for $j>10$. Then, a general regularisation is needed. However, any of these regularised expressions is $Q$-equivalent to terms with up to $r^{10}$.

\section{Spinor and pure spinor identities in $D=11$}\label{app.id}
Our convention for antisymmetrisation of indices is such that
\begin{equation}
(\g_{ab})_{\alpha\beta}=\frac{1}{2}\big[(\g_{a}\g_{b})_{\alpha\beta}-(\g_{b}\g_{a})_{\alpha\beta}\big]
\end{equation}
and the general Fierz identity is
\begin{equation}
(AB)(CD)=\sum\limits_{p=0}^5
\frac{1}{32p!}(C\g^{a_1\ldots a_p}B)(A\g_{a_p\ldots a_1}D),
\end{equation}
where the spinors have been assumed to be bosonic. With an appropriate sign dependent on the statistics of the operators added, it holds for all mixes of fermionic and bosonic operators. In specific for the pure spinor, it reduces to
\begin{equation}
(A\la)(\la B)=-\frac{1}{64}(\la\g^{ab}\la)(A\g_{ab}B)
+\frac{1}{3840}(\la\g^{abcde}\la)(A\g_{abcde}B),
\end{equation}
which results in a few useful identities for the pure spinor, some of which are
\begin{gather}\begin{aligned}\label{eq.appContr}
(\g_j\la)_\alpha(\la\g^{ij}\la)&=0\\
(\g_i\la)_\alpha(\la\g^{abcdi}\la)&=6(\g^{[ab}\la)_\alpha(\la\g^{cd]}\la)\\
(\g_{ij}\la)_\alpha(\la\g^{abcij}\la)&=-18(\g^{[a}\la)_\alpha(\la\g^{bc]}\la)\\
(\g_{ijk}\la)_\alpha(\la\g^{abijk}\la)&=-42\la_\alpha(\la\g^{ab}\la)\\
(\g_{ij}\la)_\alpha(\la\g^{abcdij}\la)&=-24(\g^{[ab}\la)_\alpha(\la\g^{cd]}\la)\\
(\g_i\la)_\alpha(\la\g^{abcdei}\la)&=\la_\alpha(\la\g^{abcde}\la)
-10(\g^{[abc}\la)_\alpha(\la\g^{de]}\la),
\end{aligned}\end{gather}
also presented in \cite{CK2}. There, we also presented the first three of the following relations, all important with respect to $L^{(n)}$:
\begin{gather}\begin{aligned}\label{eq.Leq}
(\lb\g^{[ij}\lb)(\lb\g^{kl]}r)&=0\\
(\lb\g^i{}_k\lb)(\lb\g^{jk}r)&=(\lb\g^{ij}\lb)(\lb r)\\
(\lb\g^i{}_kr)(\lb\g^{jk}r)&=(\lb\g^{ij}r)(\lb r)+\frac{1}{2}(\lb\g^{ij}\lb)(rr)\\
(\lb\g^{[ab}\lb)(\lb\g^{c]d}\lb)&=0\\
(\lb\g^{ab}\lb)(\lb\g^{cd}\lb)f_{ac}&=\frac{1}{2}(\lb\g^{ac}\lb)(\lb\g^{bd}\lb)f_{ac}\\
L^{(1)}_{ab,cd}f^{abc}&=(\lb\g_{ab}\lb)(\lb\g_{cd}r)f^{abc}.
\end{aligned}\end{gather}
For general spinor calculations,
\begin{gather}\begin{aligned}
(\g^{a_1\ldots a_p})_{\alpha\beta}&=(\g^{a_1}\g^{a_2\ldots a_p})_{\alpha\beta}-(p-1)\eta^{a_1[a_2}(\g^{a_3\ldots a_p]})_{\alpha\beta}\\
(\g^{a_1\ldots a_p})_{\alpha\beta}&=(\g^{a_1\ldots a_{p-1}}\g^{a_p})_{\alpha\beta}-(p-1)(\g^{[a_1\ldots a_{p-2}})_{\alpha\beta}\eta^{a_{p-1}]a_p}\\
(\g^{b}\g^{a_1\ldots a_p})_{\alpha\beta}&=2p\times\eta^{b[a_1}(\g^{a_2\ldots a_p]})_{\alpha\beta}+(-1)^{p}(\g^{a_1\ldots a_p}\g^{b})_{\alpha\beta}
\end{aligned}\end{gather}
can also be of use. Of course, there are plenty more relations to be deduced, but the above constitute the most important ones, used in this article.

\section{The $b$-ghost and $R^a$ in maximal supergravity}
For actual calculations, it is sometimes practical to use expressions of the $b$ and $R^a$ where gauge invariance is manifest, but not displayed in the sense of the $N$ and $N^{ab}$ operators, i.e. a reformulation of the operators in \eqref{eq.oSUGRA}:
\begin{subequations}\label{eq.oSUGRAw}
\begin{align}
&b=\frac{1}{2}\eta^{-1}(\bar{\la}\g_{ab}\bar{\la})(\la\g^{ab}\g^sD)\partial_s+\nonumber\\
&+\eta^{-2}L_{ab,cd}^{(1)}\big[(\la\g^aD)(\la\g^{bcd}D)-4(\la\g^{a[b}\la)(\la\g^{s]cd}\omega)\partial_s-6(\la\g^{[ab}\la)(\la\g^{c]}\omega)\partial^d\big]+\nonumber\\
&+\eta^{-3}L^{(2)}_{ab,cd,ef}\big[6(\la\g^{def}D)(\la\g^{[ab}\la)(\la\g^{c]}\omega)-8(\la\g^{f}D)(\la\g^{a[b}\la)(\la\g^{e]cd}\omega)\big]+\nonumber\\
&+16\eta^{-4}L^{(3)}_{ab,cd,ef,gh}(\la\g^{ab}\la)(\la\g^{ce}\la)\big\{(\la\g^{d}\omega),(\la\g^{fgh}\omega)\big\}
\end{align}
and
\begin{gather}\begin{aligned}
R^a=&\eta^{-1}(\lb\g^{ab}\lb)\partial_b-\eta^{-2}L_{(1)}^{ab,cd}(\la\g_{bcd}D)-\\
&-6\eta^{-3}L_{(2)}^{ab,cd,ef}(\la\g_{ef}\la)(\la\g_{bcd}\omega).
\end{aligned}\end{gather}
\end{subequations}
The property of $bb=0$ is easy to verify for $r^0$. It is also possible to note, with some help of \eqref{eq.Leq}, that each $\{D,D\}$ in the same relation by default yields zero, as well as that the terms with $\omega$ do not act on $\eta$ (most easily seen from \eqref{eq.oSUGRA}). Especially useful is the last relation in \eqref{eq.Leq}, due to the property of $[ab,cd]$. It is also important to remember that $L^{(n)}$ for $n>1$ contains less irreducible representations than what is allowed for by the antisymmetrisation of $n$ 2-forms:
\begin{gather}\begin{aligned}
&L^{(0)}: \mathrm{(01000)}\\
&L^{(1)}: \mathrm{(01000)}, \mathrm{(10100)}\\
&L^{(2)}: \mathrm{(00200)}, \mathrm{(02000)}, \mathrm{(10100)}, \mathrm{(20010)}\\
&L^{(3)}: \mathrm{(00200)}, \mathrm{(02000)}, \mathrm{(10110)}, \mathrm{(11100)}, \mathrm{(20010)}, \mathrm{(30002)}
\end{aligned}\end{gather}
and that additional $\lb\g^{(2)}\lb$ merely are added to the irreducible representations above, as one such is part of them all. Note also that $b_3$ does not support (20010) or (30002).

During the check of $r^1$, it is furthermore handy to observe that by \eqref{eq.Leq}
\begin{subequations}
\begin{equation}
2(\g^{mn})_{\alpha\beta}\partial^d=(\g^{d}\g^i\g^{mn})_{\alpha\beta}\partial_i
\end{equation}
when the relevant indices are contracted with $(\lb\g_{mn}\lb)(\lb\g_{cd}\lb)$ as well as that
\begin{gather}\begin{aligned}
(\lb\g_{mn}\lb)(\lb\g_{\lbracket ab}\lb)(\lb\g_{cd\rbracket}r)(\la\g^{ab}\la)(\la\g^{cd}\g^i\g^{mn})_\alpha&=\\
\frac{1}{2}(\lb\g_{mn}\lb)(\lb\g_{ab}\lb)(\lb\g_{cd}r)(\la\g^{ab}\la)(\la\g^{cd}\g^i\g^{mn})_\alpha
\end{aligned}\end{gather}
\end{subequations}
sets the $\lb^3r$ in both $[ab,cd]$ and $(ab,cd)$, i.e. to zero.

That said, we have not checked for the entire $bb=0$, but deem it extremely likely. In the general discussion it might also be interesting to note that
\begin{equation}\label{eq.L(15)}
L^{(15)}: (08000) \quad \Rightarrow \quad L^{(15)}=(\lb\g^{(2)}\lb)^8(\lb r)(rr)^7
\end{equation}
is proportional to $T$ in \eqref{eq.T}. $T$ can combine with up to $L^{(12)}$, where the combination including $r^{15}$ is in (07000), the above with one $\lb\g^{(2)}\lb$ less.

\def\prp {Phys. Rept.}
\def\prv{Phys. Rev.}
\def\prl{Phys. Rev. Lett.}
\def\prva{Phys. Rev. A}
\def\prvd{Phys. Rev. D}
\def\jhep{\mbox{J. High} Energy Phys.}
\def\pLondA{Proc. Roy. Soc. Lond. A}
\def\cqg{Class. Quantum Grav.}
{\small
\bibliography{references.bib}
\bibliographystyle{JHEP1}
}
\end{document}